# Quantum annealing for hard 2-SAT problems : Distribution and scaling of minimum energy gap and success probability

Vrinda Mehta,[1,2] Fengping Jin,[1] Hans De Raedt,[1,3] and Kristel Michielsen[1,2,4,*]

[1]*Institute for Advanced Simulation, Jülich Supercomputing Centre, Forschungszentrum Jülich, D-52425 Jülich, Germany*
[2]*RWTH Aachen University, D-52056 Aachen, Germany*
[3]*Zernike Institute for Advanced Materials, University of Groningen, Nijenborgh 4, NL-9747 AG Groningen, The Netherlands*
[4]*JARA-CSD, Jülich-Aachen Research Alliance, 52425 Jülich, Germany.*

In recent years, quantum annealing has gained the status of being a promising candidate for solving various optimization problems. Using a set of hard 2-satisfiabilty (2-SAT) problems, consisting of upto 18-variables problems, we analyze the scaling complexity of the quantum annealing algorithm and study the distributions of the minimum energy gap and the success probability. We extend the analysis of the standard quantum annealing Hamiltonian by introducing an additional term, the trigger Hamiltonian, which can be of two types : ferromagnetic and antiferromagnetic. We use these trigger Hamiltonians to study their influence on the success probability for solving the selected 2-SAT problems. We found that although the scaling of the run-time is exponential for the standard and modified quantum annealing Hamiltonians, the scaling constant in case of adding the trigger Hamiltonians can be significantly smaller. Furthermore, certain choices for the trigger Hamiltonian and annealing times can result in a better scaling than that for simulated annealing. Lastly, we also use the quantum annealers of D-Wave Systems Inc. to study their performance in solving the 2-SAT problems and compare it with the simulation results.

## I. INTRODUCTION

Quantum annealing is a metaheuristic for solving combinatorial optimization problems, which requires mapping the problem to the Ising Hamiltonian. The ground state of this so-called problem Hamiltonian encodes the solution to the optimization problem, and therefore the task of finding the solution to the optimization problem is equivalent to finding the ground state of the problem Hamiltonian. Similar to simulated annealing [1], where the search for the ground state is assisted by adding thermal fluctuations, quantum annealing makes use of quantum fluctuations so that quantum tunneling can facilitate the search for the lowest-energy configuration [2–4].

The idea of employing adiabatic quantum annealing for realizing a quantum computer devoted to solving optimization problems, emerged in the early 2000's [5, 6]. However, the notion of quantum annealing has a wider scope than adiabatic quantum computing as it also allows for non-adiabatic transitions during the evolution [7–9]. A similar algorithm that has been developed for the gate-based model of quantum computing is the quantum approximate optimization algorithm (QAOA) [10–14].

Since its conception, there has been extensive research to evaluate the efficiency of quantum annealing for solving optimization problems [15–25]. Moreover, the availability of commercial quantum annealers of D-Wave which offers annealing systems with more than 5000 qubits [26], has given an impetus to research in this direction [14, 27–39]. Much of the work has focused on investigating whether quantum annealing can deliver a speedup over the existing classical algorithms [20, 33, 40]. A related area of interest is to assess the performance of quantum annealing by studying the scaling of the computation time required to solve the optimization problem as a function of the problem size, and comparing it to the scaling of certain chosen classical algorithms [31, 33, 40–45].

In this work, we numerically investigate the scaling complexity of quantum annealing for solving 13 sets of 2-SAT problems, with the size of the problems varying from 6 to 18 variables [46]. These problems have a known ground state and have been specially designed to be hard to be solved using simulated annealing [43, 46]. Therefore, such an analysis allows us to gauge the suitability of quantum annealing for solving them.

In case of adiabatic quantum annealing, the minimum energy gap between the ground state and the first excited state of the Hamiltonian is a pivotal quantity for determining the computation time required to obtain the solution for the optimization problem [47]. Hence, we look at the scaling of the minimum energy gaps to understand how the resources required to solve these problems using quantum annealing grow as the problem size increases. For a comparison with the predictions of the adiabatic theorem, we also determine the scaling of TTS99 (time to solution) which is the run-time required to obtain the ground state of the problem Hamiltonian at least once, in multiple runs of the algorithm, with 99% certainty.

Next, to examine how modifications to the standard algorithm affects these scalings, we add a third term [9, 22, 48–52], the trigger Hamiltonian, to the standard algorithm for quantum annealing, which vanishes at the beginning and end of the annealing process. Furthermore, the trigger Hamiltonian can be of two types, the ferromagnetic trigger Hamiltonian and the antiferromagnetic trigger Hamiltonian [9, 52]. Previous findings have indicated that while the inclusion of the ferromagnetic trigger Hamiltonian mostly enlarges the minimum energy gaps, therefore promoting the chances of an adiabatic evolution, adding the antiferromagnetic trigger Hamiltonian can either increase or decrease the size of the minimum energy gaps [9, 52]. In addition, it can also modify the energy spectrum of the quantum annealing Hamiltonian significantly, for example, by increasing the number of anti-crossings between the ground state and the first excited state of the Hamiltonian, or by altering the shape of the anti-crossing, giving way to some interesting non-adiabatic mechanisms that control the evolution of the state of the system [52]. Thus, a study of the scaling of the minimum energy gaps and TTS99 with the addition of the trigger Hamiltonians facilitates a direct comparison between the adiabatic quantum annealing and the quantum annealing algorithm allowing for non-adiabatic mechanisms.

[*] k.michielsen@fz-juelich.de; Corresponding author

Finally, we also use the solvers offered by D-Wave to study the scaling of TTS99 on these systems. Although susceptible to noise and temperature effects, these systems provide the hardware for the largest quantum annealer. A comparison of the results obtained with these systems with the simulation results can give an estimate for how close these systems are to an ideal quantum annealer, as well as the dominant effects playing a role during the evolution of the state of the system.

We analyze the distribution of the minimum energy gap between the ground state and the first excited state of the quantum annealing Hamiltonian, and the success probability, as well as the scaling behavior of the minimum energy gap and the TTS99. We obtain three distinct distributions for the minimum energy gaps. In the adiabatic limit, these distributions can be extrapolated to the distribution of the success probability using the Landau-Zener theory. For certain distributions of the minimum energy gaps, the resulting distribution for the success probability is predicted to be constant. The simulation results for the success probability distribution are found to be of three kinds: bimodal, unimodal, and constant. Interestingly, the corresponding results obtained with the D-Wave annealers also show these three distributions. The scaling of the minimum energy gaps and TTS99 in the adiabatic regime, is found to be exponential in the asymptotic limit. Furthermore, we find that in the adiabatic limit, the standard quantum annealing Hamiltonian results in a worse scaling of TTS99 than brute-force search. Nevertheless, the quantum annealing algorithm with antiferromagnetic trigger Hamiltonian for short annealing times, and the new generation of the D-Wave systems, result in a better scaling than that obtained for solving these problems using simulated annealing.

The paper is organized as follows. Section II gives a brief review of the theoretical aspects of quantum annealing and related concepts. In Sec. III we describe the 2-SAT problems that have been used for this study, and briefly explain the methods used for obtaining and analyzing the results. Section IV discusses the results obtained from simulations, while in Sec. V we show the results obtained with the D-Wave systems. Finally, we conclude our observations in Sec. VI.

## II. THEORETICAL BACKGROUND

This section aims at equipping the readers with the theoretical background necessary to understand the important aspects of, or related to, quantum annealing such as the recipe for the algorithm, the adiabatic theorem, the Landau-Zener formula, and the modifications to the standard algorithm for quantum annealing.

### A. Quantum annealing and the adiabatic theorem

For employing quantum annealing to solve an optimization problem, we start the algorithm in the easy-to-prepare ground state of an initial Hamiltonian $H_I$. The system is then slowly swept towards the problem Hamiltonian $H_P$, which encodes the solution of the optimization problem in its ground state, by means of the annealing parameter $s$, defined as $s = t/T_A$, where $T_A$ is the annealing time. The time-dependent Hamiltonian has the form,

$$H(s) = A(s)H_I + B(s)H_P, \tag{1}$$

where, functions $A(s)$ and $B(s)$ control the annealing scheme, such that $A(0)/B(0) \gg 1$ and $A(1)/B(1) \ll 1$. For simplicity in our simulations work, we choose the annealing scheme to be linear, i.e., $A(s) = (1-s)$ and $B(s) = s$. Frequently, the initial Hamiltonian is chosen to be

$$H_I = -\sum_{i=1}^{N} h_i^x \sigma_i^x, \tag{2}$$

while the problem Hamiltonian is of the Ising type,

$$H_P = -\sum_{i=1}^{N} h_i^z \sigma_i^z - \sum_{\langle i,j\rangle} J_{i,j}^z \sigma_i^z \sigma_j^z, \tag{3}$$

where $\sigma_i^{x/z}$ are the Pauli matrices, $h_i^{x/z}$ are the applied fields acting along the $x/z$- directions, respectively, $J_{i,j}^z$ is the coupling between the $i^{th}$ and $j^{th}$ spins, and $\langle i, j\rangle$ denotes the set of coupled spins. $h_i^x$ is generally chosen to be 1.

The dynamics of the system described by this Hamiltonian is governed by the time dependent Schrödinger equation (TDSE). Hence, the task of finding the solution of the optimization problem is equivalent to solving the TDSE,

$$i\frac{\partial |\psi\rangle}{\partial t} = H(t)\,|\psi\rangle\,, \tag{4}$$

where we have set $\hbar = 1$, and use dimensionless quantities in our simulations.

The adiabatic theorem states that if the sweeping from the initial to the problem Hamiltonian is done slowly enough during the annealing, the system stays in the same instantaneous energy eigenstate as in which the algorithm starts [47, 53–55]. Therefore, if one starts in the ground state of the initial Hamiltonian, one reaches the state encoding the solution of the optimization problem at the end of the algorithm. This requires [47]

$$T_A \gg \max_{0\leq s\leq 1} \frac{||\langle 1(s)| \frac{dH}{ds} |0(s)\rangle ||}{\Delta(s)^2}, \tag{5}$$

where $|0(s)\rangle$ and $|1(s)\rangle$ are the ground state and first excited state of the instantaneous Hamiltonian, respectively, and $\Delta(s)$ is the energy gap between them.

As can be seen from Eq. (5), the annealing time required to reach the ground state of the problem Hamiltonian adiabatically, depends crucially on the minimum energy gap $\Delta_{min}$ between the ground state and the first excited state of the Hamiltonian, i.e., $\Delta = \min_{0\leq s\leq 1} \Delta(s)$. A Hamiltonian with a small minimum energy gap requires a long annealing time to keep the state of the system in the instantaneous ground state.

### B. The Landau-Zener theory

A test for gauging the performance of quantum annealing is to determine the probability of the final state being the required solution of the encoded problem with a given minimum energy gap and for a chosen annealing time. This probability will henceforth be referred to as the success probability.

The Landau-Zener theory describes the response of a two-level system under the action of a varying external magnetic field [56–58]. Considering a spin-1/2 particle in a time dependent magnetic field $h(t) = ct$, where $c$ is the rate of sweep, and $t$ varies from $-\infty$ to $\infty$, the Hamiltonian for the system is given by

$$H = -\Gamma\sigma_i^x - h(t)\sigma_i^z, \tag{6}$$

where $\Gamma$ sets the scale of the energy splitting between the two levels [58]. When $t$ is large and negative, and $|h(t)| \gg \Gamma$, the spin-down state is the ground state of the Hamiltonian, while for large and positive $t$, the spin-up state is the ground state of the Hamiltonian. According to the Landau-Zener theory, the success probability after the sweep is

$$p = 1 - \exp(\frac{-\pi\Delta_{min}^2}{4c}), \tag{7}$$

where $\Delta_{min} = 2\Gamma$ is the minimum energy gap between the two levels, and $c = dh/dt$. Although, in this work we deal with systems which are more complex than a simple two-level system, if the chosen annealing time is sufficiently long, the system can be well-approximated by a two-level system. Hence, the Landau-Zener formula can still be employed for determining the probability for an adiabatic evolution for such systems.

### C. Addition of the trigger Hamiltonian

In order to investigate how introducing modifications to the standard algorithm for quantum annealing can affect its performance, we add a third term [9, 52], the trigger Hamiltonian $H_T$, to the standard algorithm for quantum annealing. This term should vanish at the beginning and end of the annealing process, so that the ground states of the initial Hamiltonian and problem Hamiltonian remain unaffected. Upon choosing a linear annealing scheme, one obtains

$$H(s) = (1-s)H_I + s(1-s)H_T + sH_P. \tag{8}$$

We have chosen the trigger Hamiltonian to have the same connectivity graph as that of the problem Hamiltonian, i.e.,

$$H_T = -\sum_{\langle i,j\rangle} J_{i,j}^x \sigma_i^x \sigma_j^x. \tag{9}$$

Furthermore, the trigger Hamiltonian can be of two types, the ferromagnetic trigger Hamiltonian, with $J_{i,j}^x = 1$, and the antiferromagnetic trigger Hamiltonian with $J_{i,j}^x = -1$.

## III. PROBLEMS AND METHODS

In this section, we describe the set of problems that we want to solve using quantum annealing. Next, we explain the methods used to obtain both, numerical and D-Wave results, as well as the criteria that will be employed for the interpretation of these results.

### A. 2-SAT problems

In this work, we want to use quantum annealing for solving problems that are hard to solve with classical algorithms like simulated annealing. For accomplishing this task, we construct sets of 2-SAT problems with varying problem size, with each problem having a unique satisfying assignment (selected using the brute-force search method). Moreover, the degeneracy of the first excited state grows exponentially as the size of the problems increases [43]. Such properties make these problems difficult to solve using simulated annealing.

A 2-SAT problem consists of a cost function $F$, involving $N$ binary variables $x_i = 0, 1$, and a conjunction of $M$ clauses, such that,

$$F = (L_{1,1} \vee L_{1,2}) \wedge (L_{2,1} \vee L_{2,2}) \wedge ... \wedge (L_{M,1} \vee L_{M,2}), \tag{10}$$

where $L_{\alpha,j}$, with $\alpha = 1,...,M$ and $j = 1,2$, is a variable $x_i$ or its negation $\overline{x_i}$. A problem is considered to be satisfiable if one can find an assignment to the $x_i$'s which makes the cost function true. The problem of finding a satisfying assignment to the cost function is equivalent to finding the ground state of the Hamiltonian

$$H_{2SAT} = \sum_{\alpha=1}^{M} h_{2SAT}(\varepsilon_{\alpha,1}s_i(\alpha,1), \varepsilon_{\alpha,2}s_i(\alpha,2)), \tag{11}$$

constructed from a combination of the clauses of the 2-SAT problems, where $h_{2SAT}(s_l, s_m) = (s_l - 1)(s_m - 1)$, and $s_i(\alpha, j)$ maps the $j$th literal from the $\alpha$th clause to the Ising spin $s_i$ for $\alpha = 1,...,M$, $j = 1,2$, and $i = 1,...,N$. The variable $\varepsilon_{\alpha,j} = 1$ if $L_{\alpha,j} = x_i$, while $\varepsilon_{\alpha,j} = -1$ if $L_{\alpha,j} = \overline{x_i}$ [43, 46]. These spins are further replaced by the quantum spin operator $\sigma_i^z$ for using quantum annealing to find the minimum-energy state of the Hamiltonian.

We present results for 13 sets of 2-SAT problems, each corresponding to an $N$, with $N$ ranging from 6 to 18. The sets corresponding to small $N$ ($N < 10$) have 100 problems each, while larger sets have 1000 problems for each $N$. However, it was observed that some problems had the same graph as another problem belonging to the set, and therefore such redundancies have been removed from every set. As a result, the sets corresponding to $N < 10$ have more than 70 problems each, while the sets with $N \geq 10$ have more than 900 problems, except for sets with $N = 15$ and $N = 18$ which have 557 and 789 problems, respectively.

### B. Analysis of the numerical results

Focusing now on the numerical analysis of our study, this section describes three observables that are used as a basis for determining the complexity of quantum annealing for solving our problems in section IV. We perform this analysis for three quantum annealing algorithms, i.e., using the Hamiltonian given by Eq. (1), the one given by Eq. (8) with the ferromagnetic trigger Hamiltonian, and that given by Hamiltonian Eq. (8) with the antiferromagnetic trigger Hamiltonian. We use the three quantum annealing algorithms to solve 13 sets of 2-SAT problems, each corresponding to a different $N$.

#### 1. *Minimum energy gaps*

In the adiabatic theorem (Eq. (5)) and the Landau-Zener formula (Eq. (7)), the minimum energy gap is a decisive quantity for determining the performance of quantum annealing in the adiabatic regime. We employ the Lanczos algorithm [59] to obtain the energy spectra of the problem Hamiltonians. We investigate two aspects of the minimum energy gap: its distribution for a fixed problem size, and its scaling as a function of the problem size. We use the distribution functions given in the appendix to fit the obtained minimum energy gap distributions.

In order to inspect the scaling, we calculate the deciles for the minimum energy gaps for each quantum annealing Hamiltonian, given by Eq. (1) or Eq. (8) (using the ferromagnetic or the antiferromagnetic trigger Hamiltonian), with $N$ ranging from 6 to 18. The problems which have a minimum energy gap smaller than the first decile, $D1$, represent the hardest problems of the set, while the problems with a larger minimum energy gap than the ninth decile, $D9$, represent the easiest problems of the set. We obtain the scaling of the minimum energy gaps by fitting suitable functions to the deciles in the asymptotic limit.

According to the adiabatic theorem (Eq. (5)), this analysis can be extrapolated to provide an estimate for the annealing time required to assure an adiabatic evolution of the state of the system, for a given minimum energy gap. According to the theorem, this annealing time is inversely proportional to $\Delta_{min}^2$,

$$\ln(T_A) \propto 2\ln(\frac{1}{\Delta_{min}}). \tag{12}$$

If the correlation length $\xi = 1/\Delta_{min}$ increases exponentially as a function of the problem size, then the run-time required to keep the evolution adiabatic is also expected to grow exponentially with an exponent twice as large. Therefore, this gives an estimate for how the computation time should scale if the evolution of the state is adiabatic.

#### 2. *Success probability*

The next important observable for our analysis is the success probability, which is obtained by calculating the square of the overlap of the resulting state at the end of the annealing process with the known ground state of the problem Hamiltonian. We use the second order Suzuki-Trotter product formula algorithm [60–63] for simulating the dynamics of quantum annealing. These simulations are performed for three annealing times: $T_A = 10$, 100, and 1000, which are dimensionless since $\hbar$ has been set to 1.

As indicated by our previous study [52], for our problems, the annealing time $T_A = 10$ can be too short for the state of the system to follow the ground state adiabatically, especially for the problem sets with larger $N$. This gives way to certain non-adiabatic mechanisms to be involved in the evolution of the state of the system. On the other hand, $T_A = 1000$ was found to be sufficiently long for the success probability to follow the Landau-Zener formula for a majority of the problems. The annealing time $T_A = 100$ is the intermediate annealing time for which the difficult problems might still exhibit a non-adiabatic evolution, while the systems with larger minimum energy gaps evolve adiabatically.

We obtain the success probabilities for the three quantum annealing algorithms and for the three chosen annealing times, and plot the resulting distributions. For all the results shown, the raw success probabilities $\Gamma$ obtained from the simulations are transformed such that $\langle P_{succ}\rangle = 1/2$, where

$$P_{succ} = 1 - (1-p)^R. \tag{13}$$

The parameter $R$ can be interpreted as the number of repetitions or a scaling factor for the annealing time, required to shift the average success probability of the set to 0.5.

Looking now from a theoretical perspective, the Landau-Zener theory can provide a mapping between the minimum energy gap and the success probability [43], if the annealing time is long enough to allow for an adiabatic evolution of the system. According to the formula,

$$p = 1 - e^{-\gamma\Delta_{min}^2}, \tag{14}$$

the parameter $\gamma$ controls the speed of the sweep, which in turn is controlled by the annealing time. Thus,

$$\Delta_{min} = \gamma^{-1/2}\{-\ln(1-p)\}^{1/2}, \tag{15}$$

and the Jacobian is given by

$$||\partial\Delta_{min}/\partial p|| = \frac{1}{2}\gamma^{-1}\frac{1}{\Delta_{min}}e^{\gamma\Delta_{min}^2}. \quad (16)$$

Hence, if the probability distribution function for the minimum energy gaps, $PDF(\Delta_{min})$, is known, it is possible to obtain the success probability distribution

$$PDF(p)\,dp = C^{-1}\,PDF(\Delta_{min})\,\gamma^{-1}\frac{1}{\Delta_{min}}e^{\gamma\Delta_{min}^2}\,dp, \quad (17)$$

up to the normalization constant $C^{-1}$ [43].

If one finds the minimum energy gap distribution to follow the Weibull distribution function with $k=2$, i.e.,

$$PDF(\Delta_{min}) = a\left(\frac{\Delta_{min}}{b}\right)e^{-(\frac{\Delta_{min}}{b})^2}, \quad (18)$$

for a certain parameter $b$ and the normalization constant $a$, then substituting $PDF(\Delta_{min})$ in Eq. (17), and setting the parameter $\gamma = 1/b^2$, we obtain $PDF(P_{succ})\,dP_{succ} = PDF(p_{|_{\gamma=1/b^2}})\,dP_{succ}$ to be constant. Similarly, it is possible to obtain unimodal or bimodal distributions for the success probability distribution $PDF(P_{success})$ if one tunes the annealing time by means of $\gamma$ to the point where $\langle P_{succ}\rangle = 1/2$, depending on the distribution functions that the minimum energy gaps follow.

#### 3. *Time to Solution*

Lastly, we discuss the third metric for determining the complexity of quantum annealing, the time to solution (TTS). It is the run-time required to obtain the ground state at least once in multiple runs of the algorithm with a certain probability $P_{target}$, and is given by

$$TTS = \frac{\ln(1-P_{target})}{\ln(1-p)}T_A, \quad (19)$$

where the success probability $p$ is obtained with a single run of the algorithm with annealing time $T_A$. We define TTS99 as the run-time required to obtain at least one solution with 99% certainty, i.e., $TTS99 = TTS(P_{target} = 0.99)$.

Like in the case of the minimum energy gaps, we plot the deciles for TTS99 as a function of the problem size ($10 \leq N \leq 18$). In this case, the easier problems of the set have a run-time smaller than the first decile, while the problems having a run-time larger than the ninth decile correspond to the hard cases. For determining the scaling of TTS99 an appropriate function is fit to the deciles in the asymptotic limit.

If the annealing time is long enough for an adiabatic evolution, the success probability is mainly determined by the minimum energy gap between the ground state and the first excited state of the Hamiltonian. However, for a non-adiabatic evolution the exact energy spectrum of the problem also becomes relevant. For example, the occurrence of an even number of comparably small anti-crossings between the two lowest energy levels can be beneficial for the final success probability [52]. Such observations cannot be accounted for within the theoretical models like Landau-Zener and the adiabatic theorem. It therefore becomes interesting to compare the scaling of the theoretical run-time with the numerically obtained TTS99 scaling, in order to understand the ways in which the non-adiabatic mechanisms can contribute to the scaling.

### C. Analysis of the D-Wave results

This section briefly describes how the D-Wave quantum annealer is employed for solving the sets of problems, and how the obtained results (shown in Sec. V) are analyzed. Currently, D-Wave offers systems with two types of quantum chip topolgies. The first is the Chimera topology available on the DW_2000Q_6 system (DW2000Q), while the more recent and better connected one is the Pegasus topology available on the Advantage_system1.1 system (DWAdv) [26]. For finding the ground state of our transverse-field Ising Hamiltonian using the D-Wave annealer, the problem Hamiltonian needs to be mapped to the working graph of the system. However, it is possible that a Hamiltonian cannot be directly embedded on to the system, and requires two or more physical qubits to be grouped together to represent a logical qubit of the Hamiltonian instead. Since the Pegasus topology has a higher connectivity compared to that of the Chimera topology, most of the problems could be directly embedded to DWAdv, whereas, approximately, only half of the cases had a direct mapping on DW2000Q. For example, for the set corresponding to $N = 17$, only 489 out of 913 problems have a direct mapping on DW2000Q, but this number increases to 854 on DWAdv.

We employ both DW2000Q and DWAdv systems for solving nine sets of problems ($10 \leq N \leq 18$), and choose annealing times of 4, 20, and 100 $\mu s$. The success probabilities in this case are determined by finding the ratio of the outcomes with the correct ground state energy, to the total number of samples. For annealing times of 4 $\mu s$ and 20 $\mu s$, the total number of samples is chosen to be 10000, while there are 2000 samples for $T_A = 100\ \mu s$. We gauge the performance of the D-Wave systems by observing the success probability distributions by mapping the raw success probability $p$ to success probability $P_{succ}$ using Eq. (13), and the scaling of TTS99, and compare it to the results obtained from the simulations.

## IV. NUMERICAL RESULTS

In this section, we present the simulation results, for the three previously described observables, which can help us to understand the complexity of the quantum annealing algorithm in solving the selected problems. The following sections discuss these observables one by one.

### A. Minimum energy gap analysis

We begin by addressing the results obtained for the static quantifier of quantum annealing, i.e., the minimum energy gap where we have further separated the distributions of the minimum energy gaps from the scaling results as a function of the problem size.

#### 1. Minimum energy gap distributions

Figure 1 shows the median-normalized minimum energy gap distributions for the standard quantum annealing Hamiltonian given by Eq. (1), for the three problem sets with the largest problem sizes, i.e., $N$ = 16, 17, and 18. It can be seen that the distribution agrees well with the Fréchet function (see Appendix ) and that the value of the variable $k$ (see Eq. (A.1)) decreases as the problem size becomes larger, tending towards the value of 1 as $N$ increases.

The minimum energy gap distributions for the quantum annealing Hamiltonian given by Eq. (8) with the ferromagnetic trigger Hamiltonian, as shown in Fig. 2, differ significantly from the distributions for the standard quantum annealing Hamiltonian given by Eq. (1). Previous studies have suggested that the addition of the ferromagnetic trigger Hamiltonian to the standard Hamiltonian for quantum annealing can result in an enlargement of the minimum energy gaps for almost all the problems belonging to the studied sets [52]. The distribution therefore is very different from that of the standard quantum annealing Hamiltonian, and is similar to the normal distribution.

For obtaining the fits for the minimum energy gap distribution, we fit translated-Weibull functions (Eq. (A.3)) to the distribution of correlation length $\xi = 1/\Delta_{min}$. In this case, the parameter $\mu$ in Eq. (A.3) can be interpreted as the shift due to adding the ferromagnetic trigger Hamiltonian. By setting the parameters $\mu$, $b$, and $k$ obtained from fitting the translated-Weibull function to the median-normalized correlation length distribution, we fit the transformed-translated-Weibull function (Eq. (A.4)) to the corresponding median-normalized minimum energy gap distribution by the parameter $a$. The resulting fits match the form of the distribution well.

Finally, the median-normalized minimum energy gap distributions for the quantum annealing Hamiltonian given by Eq. (8) with the antiferromagnetic trigger Hamiltonian are shown in Fig. 3. For this case, exponentially decaying functions match the distribution well. This function is equivalent to the Weibull distribution with $k = 1$.

At this point, it should be noted that three distinct minimum energy gap distributions are obtained for the three quantum annealing Hamiltonians discussed. Therefore, according to Eq. (17), we can expect to obtain three kinds of success probability distributions for the three quantum annealing Hamiltonians.

#### 2. Scaling of the minimum energy gaps

In order to investigate the complexity of the quantum annealing algorithm in solving the set of problems considered in this work, we now turn to the scaling aspect of the minimum energy gaps as a function of the problem size. Figure 4 shows the odd deciles for the minimum energy gaps for the three types of quantum annealing Hamiltonians examined in this study. It can be observed that for all the three cases, the decile values decrease exponentially as a function of the problem size in the asymptotic limit. Hence, we use exponents obtained by fitting the exponential functions $\Delta_{min} = De^{-r_\Delta N}$ to the deciles, for determining the scaling of the minimum energy gaps. For the quantum annealing Hamiltonians given by Eq. (1), and Eq. (8) with the ferromagnetic trigger Hamiltonian, the fitting is done for the sets with $N \geq 15$, whereas for the Hamiltonian given by Eq. (8) with the antiferromagnetic trigger Hamiltonian, the exponential function is used to fit the sets with $N \geq 12$.

The exponential vanishing of the minimum energy gaps with the increasing size of the problems in the asymptotic limit, confirms the hardness of these problems. From Fig. 4, it can also be seen that while for the standard quantum annealing Hamiltonians given by Eq. (1) and Eq. (8) with the ferromagnetic trigger Hamiltonian, the exponent $r_\Delta$ grows monotonically as one goes from $D1$ to $D9$, for the Hamiltonian given by Eq. (8) with the antiferromagnetic trigger Hamiltonian, the exponents stay similar for all the deciles. The median minimum energy gap, given by $D5$, scales with a rate of $r_\Delta = -0.541$ for the standard quantum annealing Hamiltonian given by Eq. (1), whereas $r_\Delta = -0.230$ and $r_\Delta = -0.316$ for the quantum annealing Hamiltonian given by Eq. (8) with the ferromagnetic trigger Hamiltonian and the antiferromagnetic trigger Hamiltonian, respectively. Therefore, the addition of both the triggers, improves the scaling of the median minimum energy gap, despite of the finding that adding the antiferromagnetic trigger can either enlarge or reduce the minimum energy gaps between the ground state and first excited states of a Hamiltonian [9, 52].

Using Eq.(12), we extend this analysis to obtain the scaling of the theoretical run-time (TR), which provides an estimate for how the computation time required for ensuring an adiabatic evolution of the state of the system grows as the size of the problem increases. We therefore expect that the median run-time should scale with a rate of $r_{TR} = 1.082$ for the quantum annealing algorithm using the Hamiltonian Eq. (1), if one were to fit functions of the form $T_A = D\exp(r_{TR}N)$ to the corresponding plots. Similarly, for the quantum annealing algorithm using the Hamiltonian Eq. (8) with the ferromagnetic trigger Hamiltonian, the median run-time is expected to grow with an exponent $r_{TR} = 0.460$, while for the algorithm using the Hamiltonian Eq. (8) with the antiferromagnetic trigger Hamiltonian, $r_{TR}$ is expected to be 0.632 in the median.

Since a brute-force search for the ground state of the problem Hamiltonian scales as $2^N$ with the Hilbert space, and therefore with an exponent of $\ln(2) = 0.693$, even a simple random generation of the eigenstates can yield the ground state of the problem Hamiltonian faster than the standard algorithm for quantum annealing, i.e., with Hamiltonian Eq. (1). However, the expected run-time improves upon adding both the trigger Hamiltonians, not only in comparison to the standard quantum annealing algorithm, but also compared to the brute-force search, especially for the quantum annealing algorithm with the ferromagnetic trigger Hamiltonian.

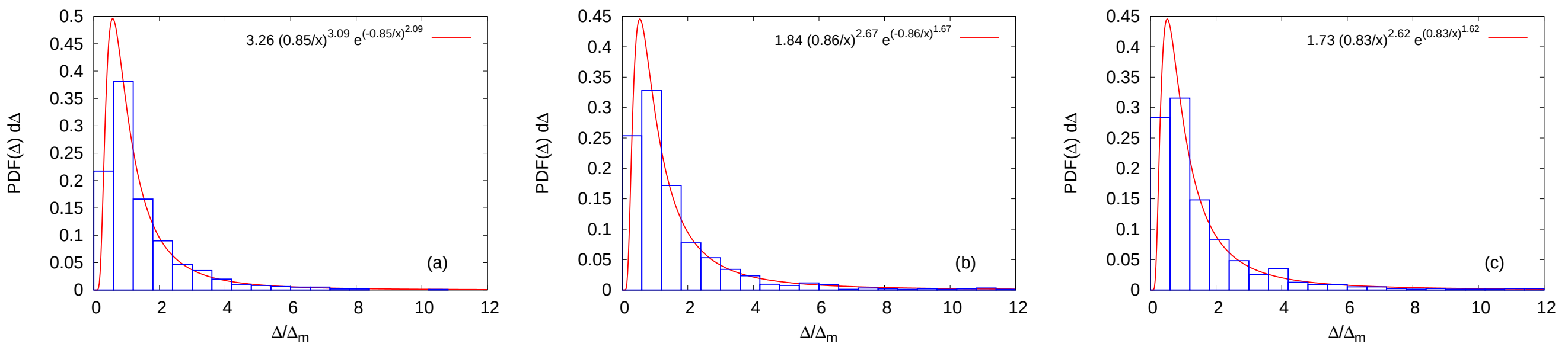

FIG. 1: (color online) Median-normalized minimum energy gap distributions $PDF(x)dx$ for the standard quantum annealing Hamiltonian given by Eq. (1), for sets with $N = 16$ (a), $N = 17$ (b), and $N = 18$ (c), where $x = \Delta/\Delta_m$ and $\Delta_m$ is the median minimum energy gap of the set.

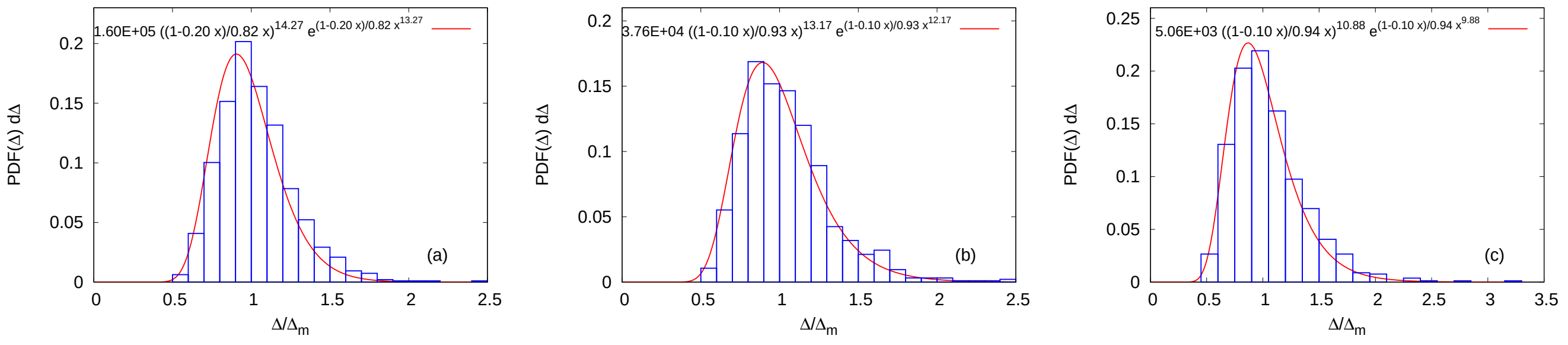

FIG. 2: (color online) Median-normalized minimum energy gap distributions $PDF(x)dx$ for the quantum annealing Hamiltonian given by Eq. (8) with the ferromagnetic trigger Hamiltonian, for sets with $N = 16$ (a), $N = 17$ (b), and $N = 18$ (c), where $x = \Delta/\Delta_m$ and $\Delta_m$ is the median minimum energy gap of the set.

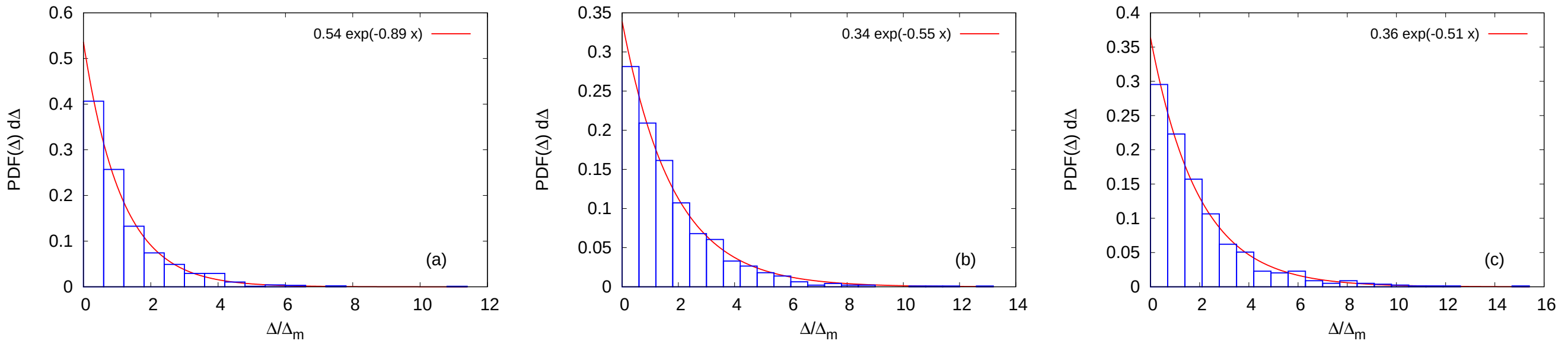

FIG. 3: (color online) Median-normalized minimum energy gap distributions $PDF(x)dx$ for the quantum annealing Hamiltonian given by Eq. (8) with the antiferromagnetic trigger Hamiltonian, for sets with $N = 16$ (a), $N = 17$ (b), and $N = 18$ (c), where $x = \Delta/\Delta_m$ and $\Delta_m$ is the median minimum energy gap of the set.

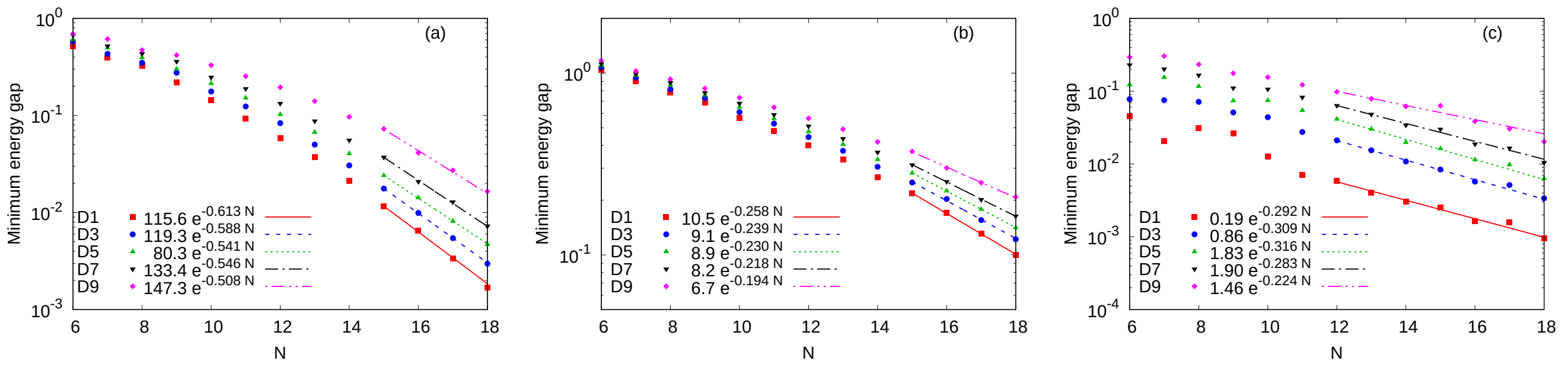

FIG. 4: (color online) Deciles for the minimum energy gaps for the quantum annealing Hamiltonian given by Eq. (1) (a), quantum annealing Hamiltonian given by Eq. (8) with the ferromagnetic trigger Hamiltonian (b), and quantum annealing Hamiltonian given by Eq. (8) with the antiferromagnetic trigger Hamiltonian (c).

### B. Success probability distributions

Having discussed the static quantifier of the quantum annealing search complexity for our problems, we now move on to address a dynamic quantifier - the success probability obtained using the three quantum annealing Hamiltonians considered in this work, for three annealing times.

Figure 5 shows the mapped success probability distributions obtained, according to Eq. (13), from the quantum annealing algorithm with Hamiltonian Eq. (1), for problem sets with $N = 17, 18$ and annealing times of $T_A = 10, 100, 1000$. Using the energy scale of a D-Wave annealer this translates to annealing times of 0.5 ns, 5 ns, and 50 ns, respectively. It can be observed that the resulting distribution is bimodal for all the annealing times, and both the problem sets. Although not for 2-SAT problems, similar results have been obtained for the success probability distribution for solving spin glass problems using simulated quantum annealing and a 108 qubit D-Wave One system [31].

A similar treatment of the success probability distributions obtained for the quantum annealing algorithm using the Hamiltonian Eq. (8) with the ferromagnetic trigger Hamiltonian, however, results in unimodal distributions, as shown in Fig. 6. This is in contrast with the observations in [31], where a unimodal distribution is only obtained for the success probability of solving spin glass problems using simulated annealing.

Interestingly, the mapped success probability distributions for the quantum annealing algorithm using the Hamiltonian Eq. (8) with the antiferromagnetic trigger Hamiltonian result in two kinds of distributions, depending on the chosen annealing time. It can be seen from Fig. 7, that in both the sets ($N = 17$ and $N = 18$) the distributions corresponding to $T_A = 10$ seem rather constant, whereas on increasing the annealing time, like in [31], the distributions show bimodality. It should be noted that this instance does not correspond to the case for which the Landau-Zener theory predicts a constant distribution (section III B 2). The difference originates from the fact that the theoretical mapping between the distributions for the minimum energy gap and the success probability utilizes the Landau-Zener formula, which only holds in the adiabatic limit, i.e., for long annealing times. This, however, is not the case here as for $T_A = 10$, especially upon adding the antiferromagnetic trigger Hamiltonian, non-adiabatic mechanisms can additionally be responsible for improving the success probability, as can be confirmed from [52].

Nevertheless, the success probability distributions from the simulations for the dynamics of quantum annealing can result in three types of distributions, i.e., unimodal, bimodal and constant distributions.

### C. Scaling of TTS99

In this subsection, we present the scaling results for another dynamic quantifier of the quantum annealing search complexity, the time to solution (TTS) with 99% certainty, obtained by using Eq. (19). Figure 8 shows the deciles for TTS99 obtained from the success probabilities for the quantum annealing algorithm using the Hamiltonian Eq. (1) with $10 \leq N \leq 18$ and for $T_A = 10, 100, 1000$.

As a first observation, it should be noted that as expected, the run-time for all the deciles increases exponentially as the problem size increases in the asymptotic limit, and thus fitting function $TTS99 = D\exp(r_{TTS99}N)$ has been used here (see Fig. 8). Secondly, the median TTS99 scales with a rate $r_{TTS99} = 0.530$ for $T_A = 10$, but increases to $r_{TTS99} = 1.170$ and $r_{TTS99} = 1.205$ as the annealing time is increased to 100 and 1000, respectively. This is an interesting observation as it suggests that for long annealing times, the run-time required to obtain at least one solution with 99% probability scales with a similar exponent as predicted by the adiabatic theorem for the run-time as a function of the minimum energy gaps ($r_{TR} = 1.082$ for the standard quantum annealing algorithm, as given in IV A 2), and thus worse than the brute force search. For the short annealing time, however, the exponent is significantly smaller than the theoretical prediction, and is also better than the brute force search. This trend is also observed for the other two quantum annealing algorithms using the Hamiltonian Eq. (8), with the ferromagnetic (Fig. 9) and the antiferromagnetic trigger Hamiltonian (Fig. 10), and can be explained as follows. The annealing time $T_A = 10$ is not long enough for the state of the system to evolve adiabatically, and hence the non-adiabatic mechanism of fast annealing can play a role in improving the success probabilities [52], making the exponent of the scaling of the median TTS99 smaller. Upon increasing the annealing time, the state of the system follows the adiabatic theorem, and therefore the dynamically obtained run-times agree well with the theoretically predicted run-times.

Additionally, by comparison of the algorithms with the three Hamiltonians, it can be seen that for the short annealing time of $T_A = 10$, the quantum annealing algorithm using the Hamiltonian Eq. (8) with the antiferromagnetic trigger Hamiltonian results in the best scaling ($r_{TTS99} = 0.277$ for the median TTS99). This is the regime where non-adiabatic mechanisms can play a significant role in the evolution of the system, and the addition of the antiferromagnetic trigger enhances the non-adiabatic effects, thus improving the scaling. On the other hand, in the long annealing time limit, where the evolution is mainly adiabatic, quantum annealing using the Hamiltonian Eq. (8) with the ferromagnetic trigger Hamiltonian shows the best scaling ($r_{TTS99} = 0.478$ for $T_A = 1000$ for median TTS99), consistent with the scaling obtained for the minimum energy gaps.

We can also observe, that even for longer annealing times,

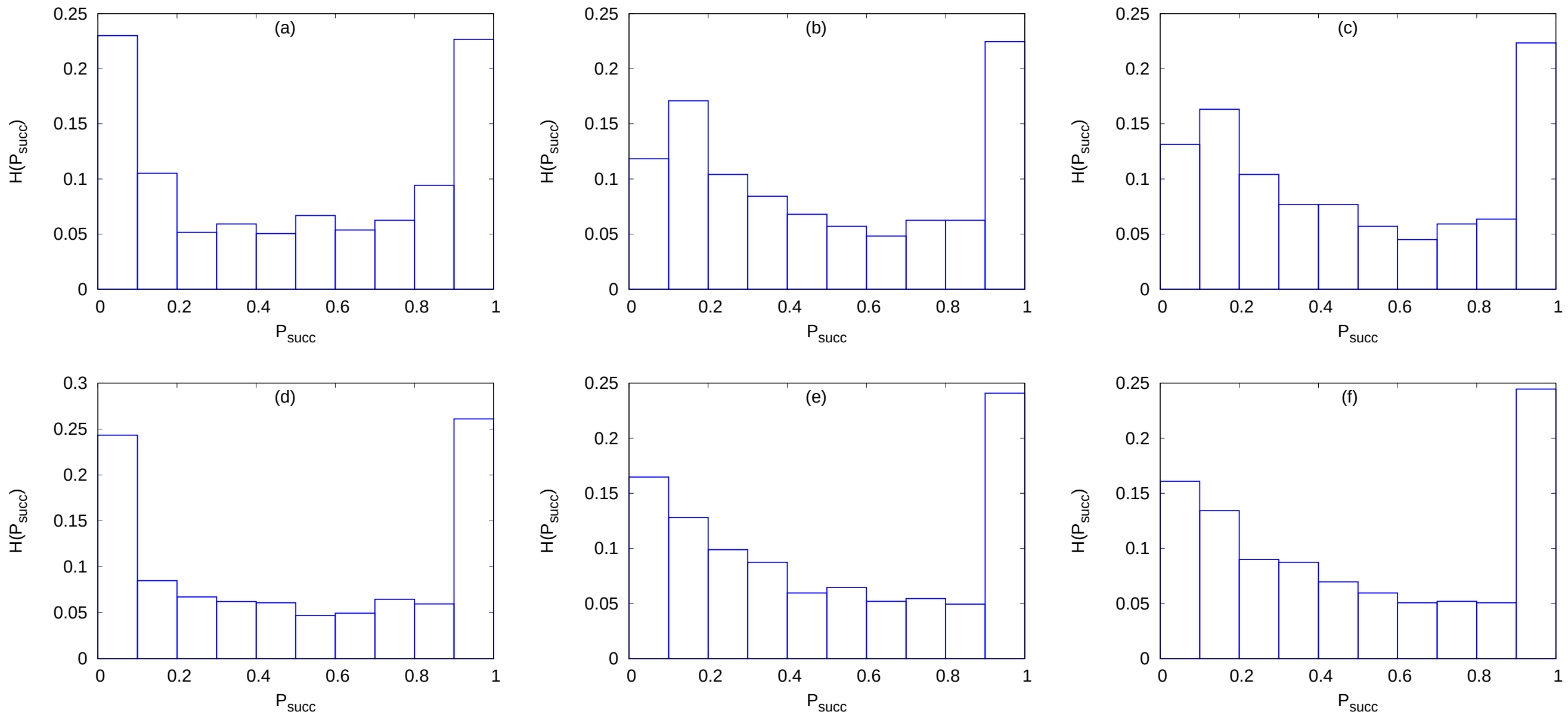

FIG. 5: (color online) Success probability distributions obtained for the quantum annealing algorithm using the Hamiltonian Eq. (1), for problem sets with $N = 17$ (a), (b), (c) and $N = 18$ (d), (e), (f) for $T_A = 10$ (a), (d), $T_A = 100$ (b), (e), and $T_A = 1000$ (c), (f).

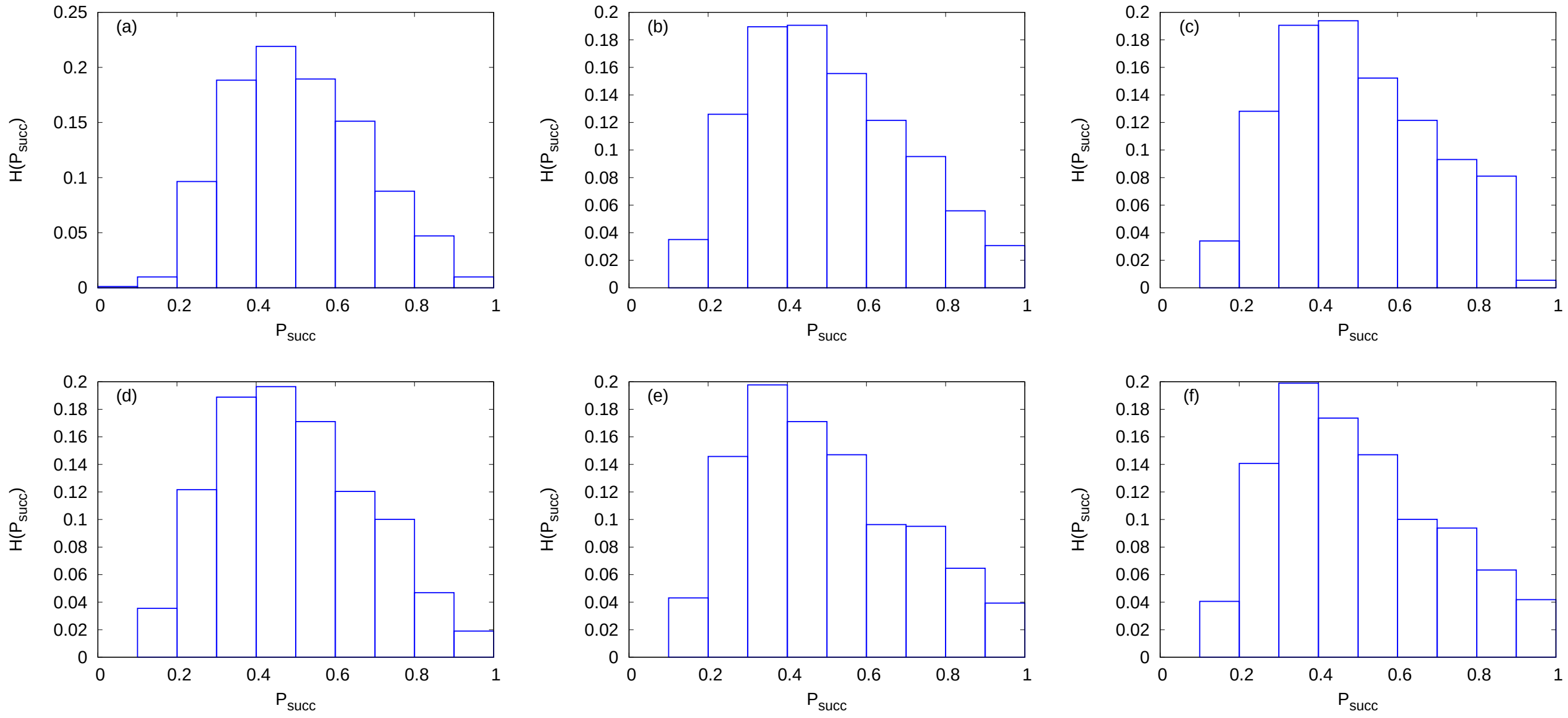

FIG. 6: (color online) Success probability distributions obtained for the quantum annealing Hamiltonian given by Eq. (8) with the ferromagnetic trigger Hamiltonian, for problem sets with $N = 17$ (a), (b), (c) and $N = 18$ (d), (e), (f) for $T_A = 10$ (a), (d), $T_A = 100$ (b), (e), and $T_A = 1000$ (c), (f).

the exponent $r_{TTS99}$ slightly varies from the theoretically predicted $r_{TR}$. For understanding the reason for such a deviation, it can be noted from the success probability versus minimum energy plots in [52] that for our set of problems, the exponents of the minimum energy gap in the Landau-Zener formula are slightly different from 2, for the three quantum annealing Hamiltonians.

## V. D-WAVE RESULTS

Finally, in this section, we perform a similar analysis with the data obtained with the two D-Wave systems as was performed with the numerical results, by examining the success probability distributions and the scaling of TTS99.

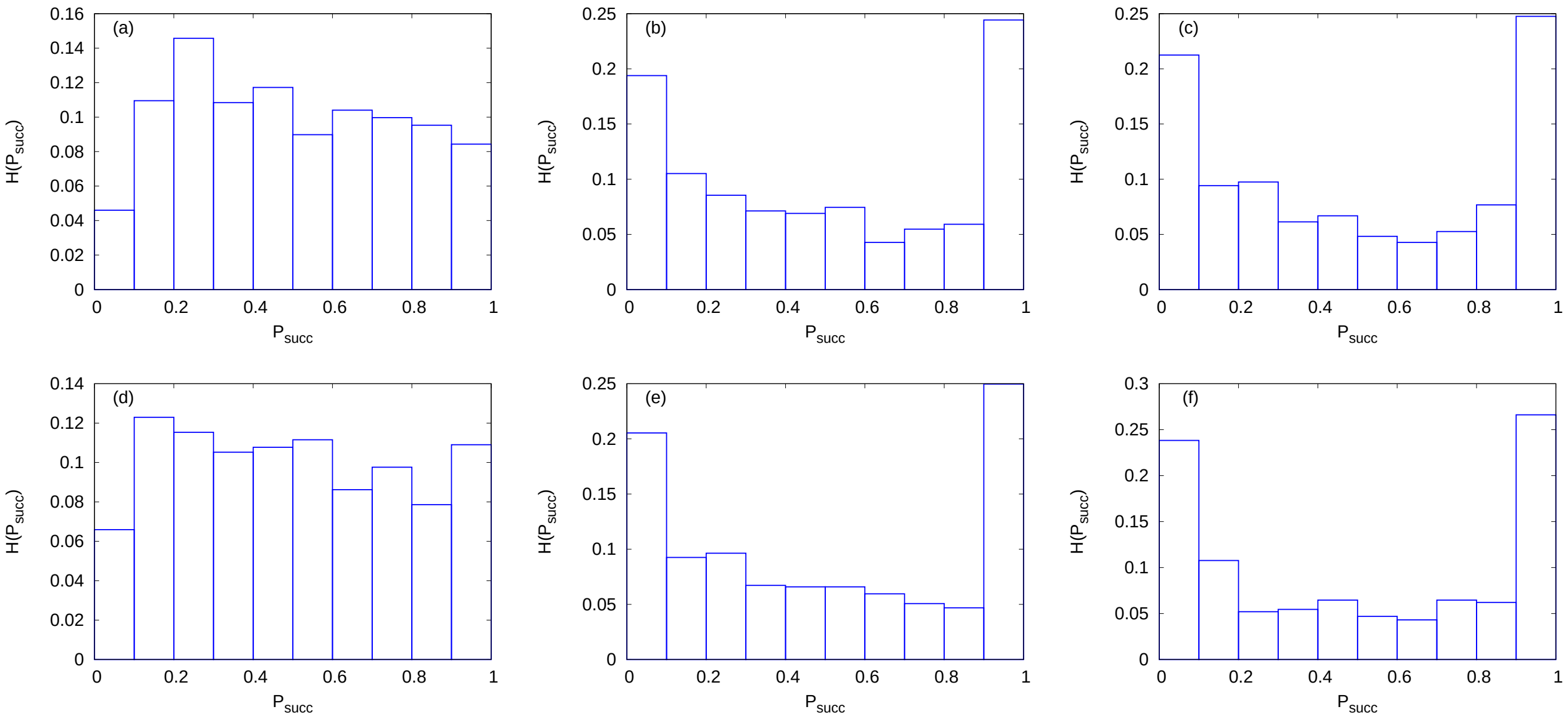

FIG. 7: (color online) Success probability distributions obtained using the quantum annealing Hamiltonian Eq. (8) with the antiferromagnetic trigger Hamiltonian, for problem sets with $N = 17$ (a), (b), (c) and $N = 18$ (d), (e), (f) for $T_A = 10$ (a), (d), $T_A = 100$ (b), (e), and $T_A = 1000$ (c), (f).

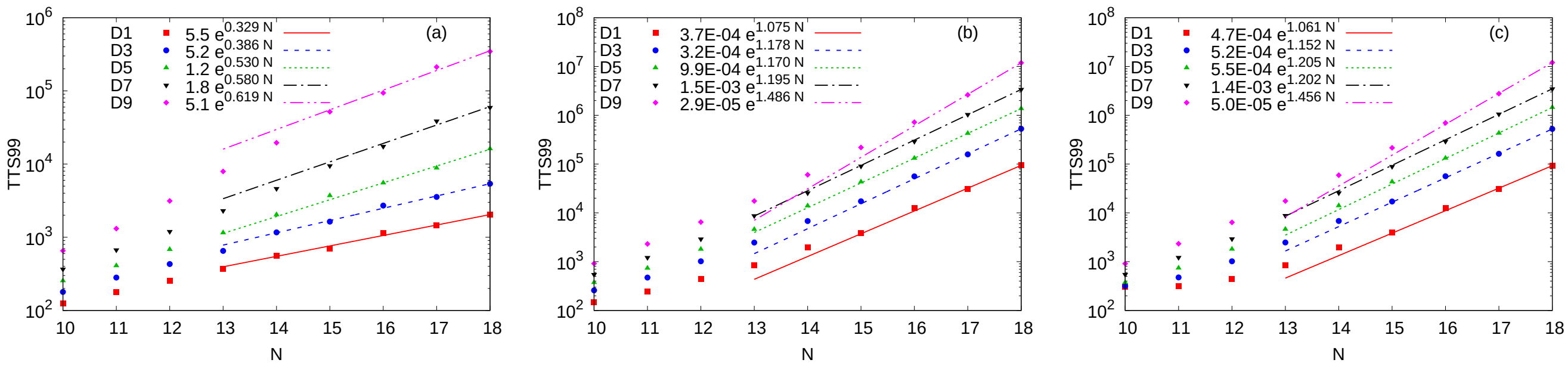

FIG. 8: (color online) Deciles for TTS99 for the quantum annealing Hamiltonian given by Eq. (1), for $T_A = 10$ (a), $T_A = 100$ (b), and $T_A = 1000$ (c).

### A. Success probability distributions

Using Eq. (13), we map the raw success probabilities $p$ such that $\langle P_{succ} \rangle = 1/2$, as was done for the numerical results. Figure 11 shows the mapped success probability distributions for sets with $N = 17, 18$ and annealing times $T_A = 4, 20, 100\ \mu s$ using DW2000Q. It can be seen that for all annealing times, and both the problem sets, the distribution is always bimodal, as was the numerically obtained success probability distribution using the standard quantum Hamiltonian given in Eq. (1). Interestingly, the mapped success probability distributions for the same problem sets and annealing times, but obtained with DWAdv are significantly different, as shown in given in Fig. 12. These distributions are not bimodal, but instead are closer to unimodal and constant distributions. Such distributions were also obtained from the simulations of the dynamics of quantum annealing using the Hamiltonian Eq. (8) with the ferromagnetic and the antiferromagnetic trigger. However, the similar results from the simulations correspond to different quantum annealing Hamiltonians than the standard Hamiltonian implemented by D-Wave. Therefore, unlike the simulations, the D-Wave systems are not ideal and many other effects, like temperature and noise, play a major role during the evolution of the system.

### B. Scaling of TTS99

We discuss the scaling of the deciles for TTS99, as was done in Sec. IV C, from the success probabilities obtained with the two D-Wave systems. In calculating the deciles for TTS99, we have omitted the problems which could not be directly embedded onto the D-Wave system in order to have a fair comparison with the scaling results obtained from the simulations.

We begin with Fig. 13 showing the TTS99 deciles obtained from the success probabilities using DW2000Q.

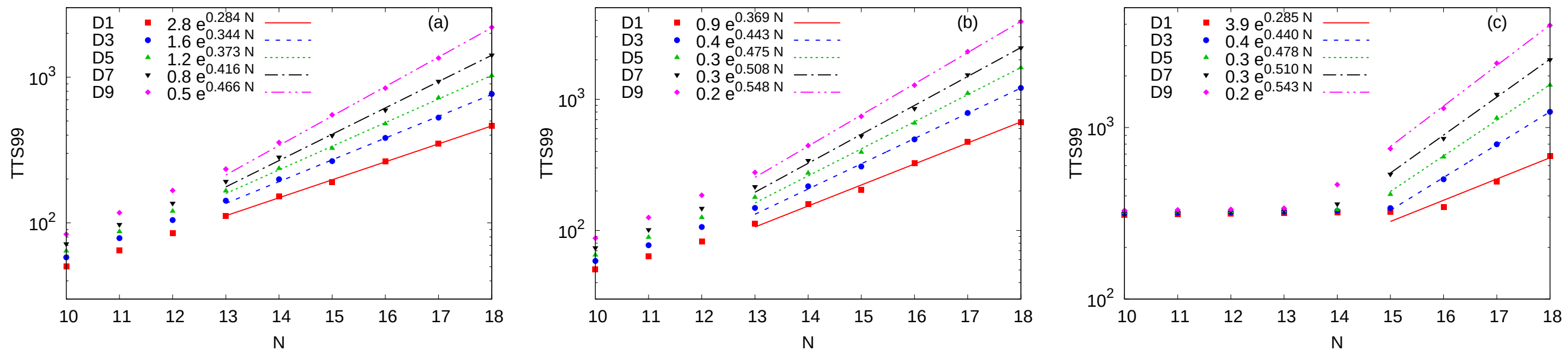

FIG. 9: (color online) Deciles for TTS99 for the quantum annealing Hamiltonian given by Eq. (8) with the ferromagnetic trigger Hamiltonian, for $T_A = 10$ (a), $T_A = 100$ (b), and $T_A = 1000$ (c).

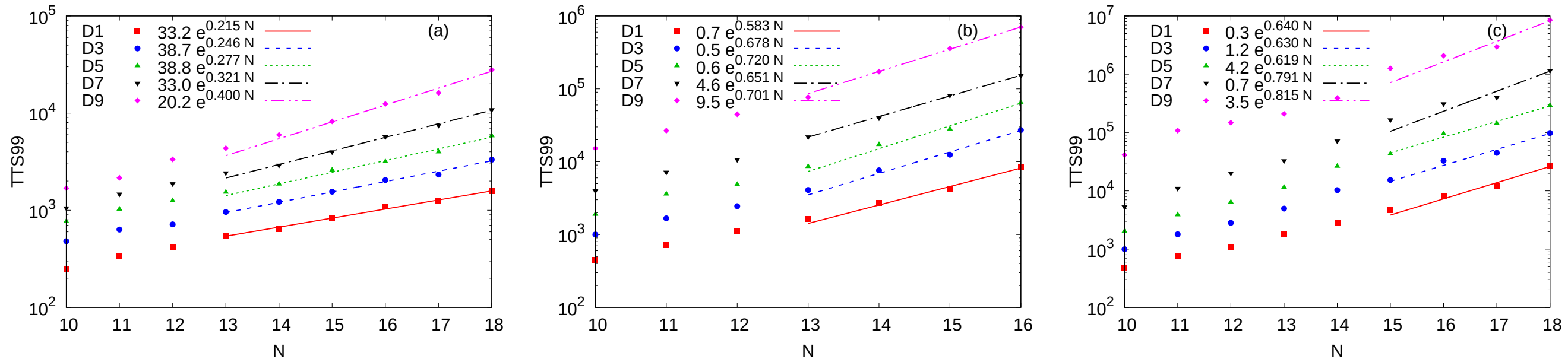

FIG. 10: (color online) Deciles for TTS99 for the quantum annealing Hamiltonian given by Eq. (8) with the antiferromagnetic trigger Hamiltonian, for $T_A = 10$ (a), $T_A = 100$ (b), and $T_A = 1000$ (c).

In this case too, in the asymptotic limit the TTS99 deciles are found to increase exponentially as the problem size increases, and hence are fit to functions $TTS99 = D\exp(r_{TTS99}N)$. The exponents $r_{TTS99}$ obtained from the fitting for the scaling of the median are 0.674, 0.511 and 0.531 for annealing times of 4, 20 and 100 $\mu s$, respectively. These values are better than the theoretical estimate for the run-time obtained from the adiabatic theorem ($r_{TR} = 1.086$ from section IV A 2), and than the exponent for the brute force search. Moreover, for the annealing times 20 $\mu s$ and 100 $\mu s$, the scaling exponents are similar to the scaling rate of the numerical TTS99 for the quantum annealing algorithm using the Hamiltonian Eq. (1) in the fast annealing limit ($r_{TTS99} = 0.530$ for $T_A = 10$ in section IV C). This behaviour can be explained on the basis of the noise present in the D-Wave system, which can give way to several non-adiabatic mechanisms improving the success probability.

A similar treatment of the deciles for TTS99 obtained from the success probabilities using DWAdv, results in an even better scaling, as shown in Fig. 14. In this case, $r_{TTS99}$ obtained from the fits to the median TTS99 is found to be 0.393, 0.325 and 0.302 for the annealing times of 4, 20 and 100 $\mu s$. Although there is a need for further research, the topological differences between the connectivity of the qubits in the two D-Wave systems can offer a plausible explanation for the dissimilarity between the scaling results from the two systems. While a qubit in DW2000Q has a connectivity of 6, in DWAdv, each qubit is connected to 15 other qubits. This can lead to additional noise being present in DWAdv, which can contribute towards a better scaling performance.

Another distinguishing factor shown by the scaling results from the D-Wave systems compared to the simulation results is that the scaling exponents $r_{TTS99}$ become approximately smaller as the annealing time is increased on both the D-Wave systems. A probable explanation for such a behavior is a more significant effect of the noise present in the systems for longer annealing times, as the system is in the quasistatic limit [64]. Moreover, unlike the annealing schedule used in the simulations, the D-Wave annealers do not use linear functions for $A(s)$ and $B(s)$.

## VI. CONCLUSION

The goal of this work was to assess the performance of quantum annealing for solving 13 sets of hard 2-SAT problems corresponding to different problem sizes, using both, simulations and the D-Wave quantum annealer. In addition to the standard Hamiltonian used for quantum annealing, we studied the performance of the algorithm for introducing two variations to the standard Hamiltonian in our simulations, by adding the ferromagnetic or the antiferromagnetic trigger Hamiltonian. The performance of the algorithm is determined by studying the distributions and/or scalings of three observables: the minimum energy gap between the ground state and the first excited state of the Hamiltonian, the success probability, and the time to solution (TTS). The smaller sets of the studied problems ($6 \leq N < 10$) have more than 70 problems each, while the larger sets ($10 \leq N \leq 18$) have more than 900 problems, except for the sets with $N = 15$ and $N = 18$.

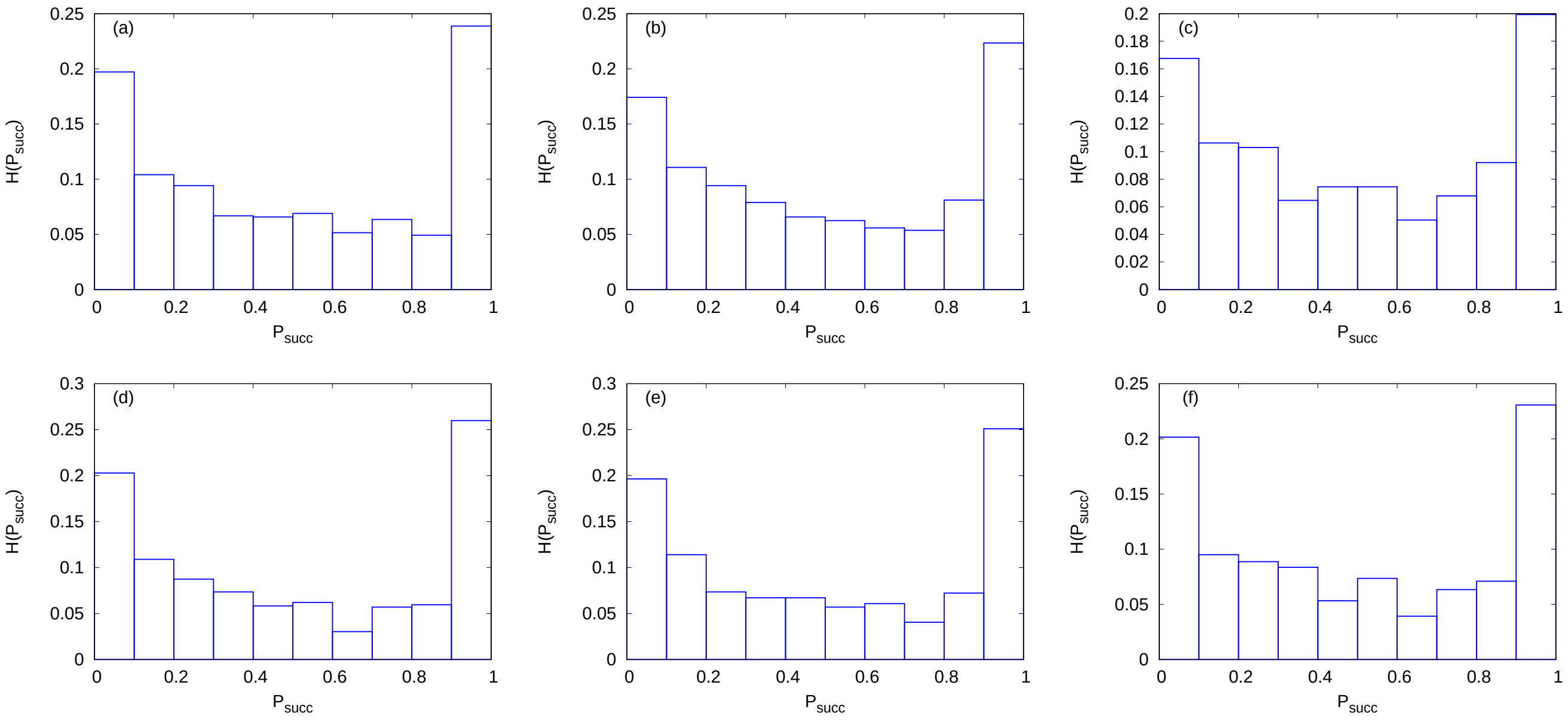

FIG. 11: (color online) Success probability distributions obtained using DW2000Q for sets with $N = 17$ (a), (b), (c) and $N = 18$ (d), (e), (f) for $T_A = 4\ \mu s$ (a), (d), $T_A = 20\ \mu s$ (b), (e), and $T_A = 100\ \mu s$ (c), (f).

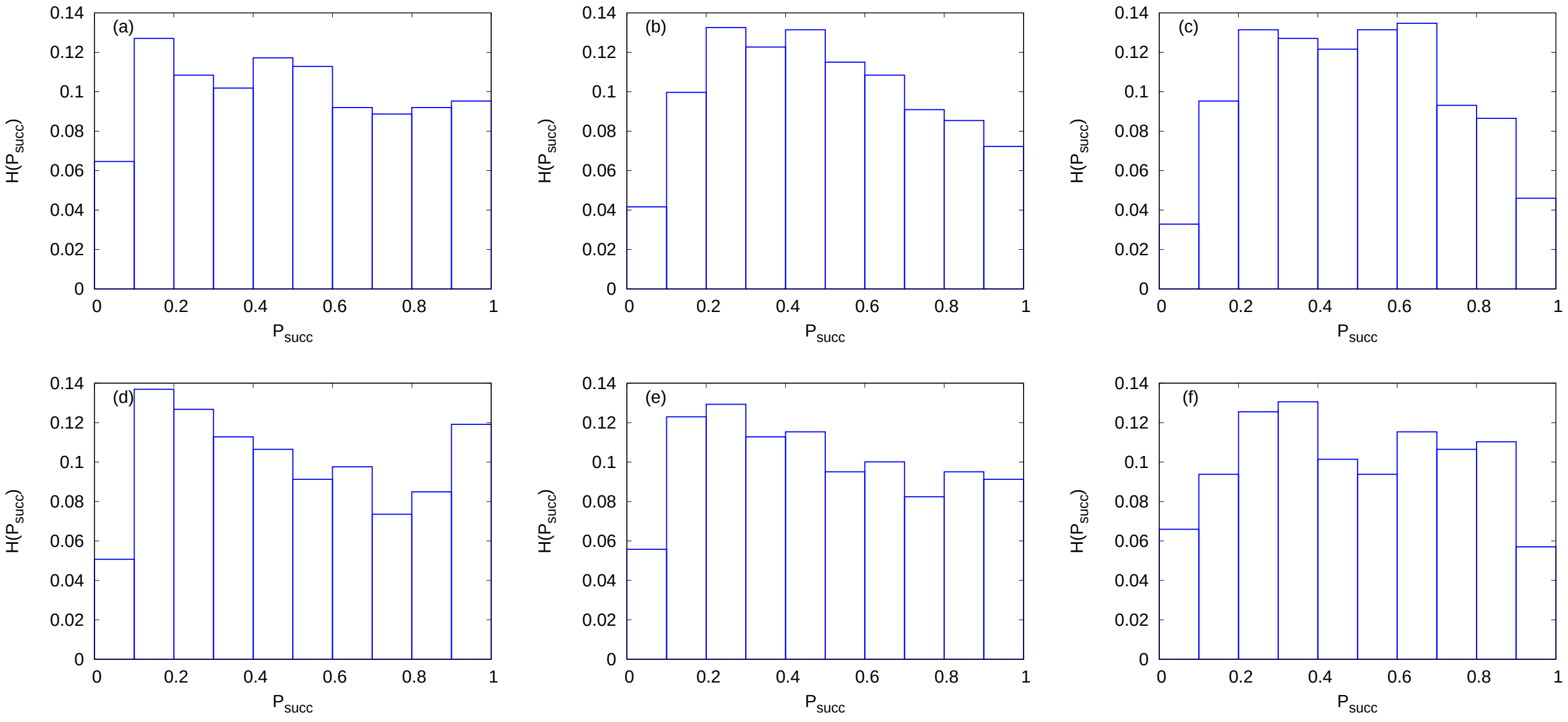

FIG. 12: (color online) Success probability distributions obtained using DWAdv for sets with $N = 17$ (a), (b), (c) and $N = 18$ (d), (e), (f) for $T_A = 4\ \mu s$ (a), (d), $T_A = 20\ \mu s$ (b), (e), and $T_A = 100\ \mu s$ (c), (f).

The minimum energy gap is a vital quantity for determining the annealing time required to reach the ground state of the problem Hamiltonian adiabatically or equivalently, the probability of finding the ground state of the problem Hamiltonian for a chosen annealing time. We find, that our problems exhibit different minimum energy gap distributions depending on the Hamiltonian chosen for the quantum annealing algorithm. The minimum energy gap distributions using the standard Hamiltonian for quantum annealing resemble the Fréchet distributions, while the distributions for the quantum annealing Hamiltonian upon adding the ferromagnetic and antiferromagnetic trigger Hamiltonian result in the transformed-translated Weibull distributions and exponential distributions, respectively. Using the Landau-Zener theorem it is possible to transform the distribution of the minimum energy gaps to that of the success probability for a sufficiently long annealing time.

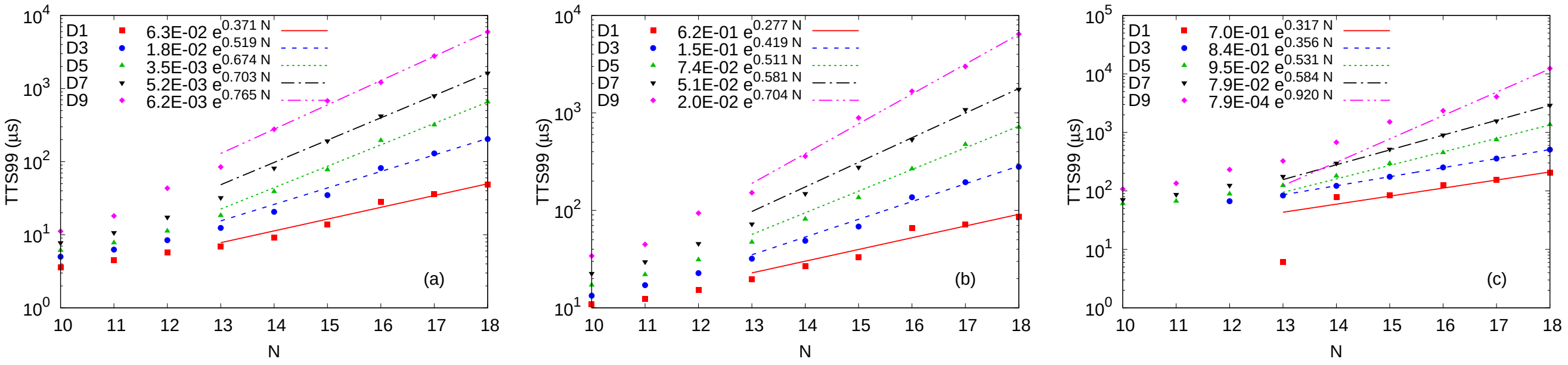

FIG. 13: (color online) Deciles for TTS99 for the cases with a native embedding on DW2000Q for $T_A = 4\ \mu s$ (a), $T_A = 20\ \mu s$ (b), and $T_A = 100\ \mu s$ (c).

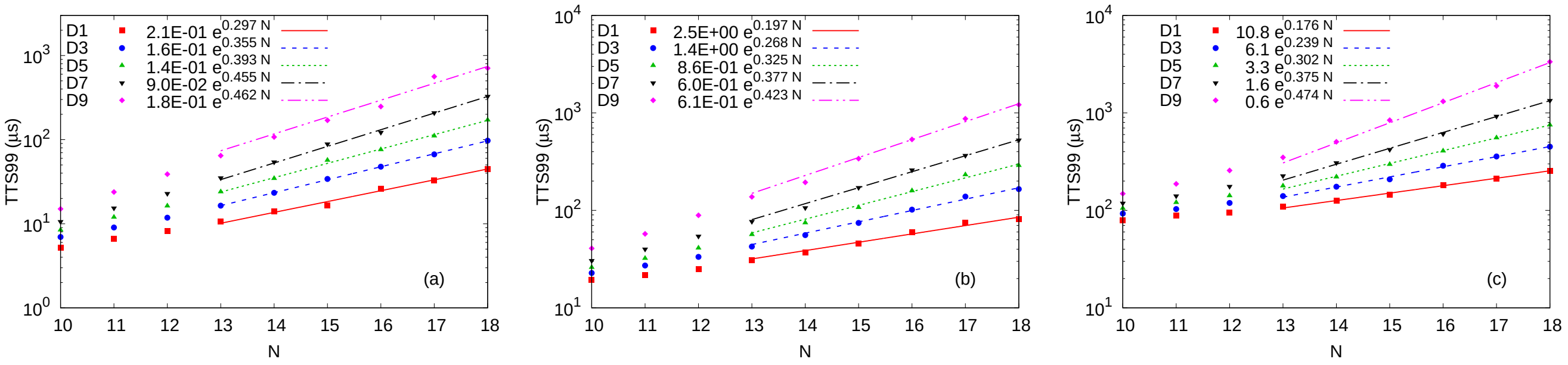

FIG. 14: (color online) Deciles for TTS99 for the cases with a native embedding on DWAdv for $T_A = 4\ \mu s$ (a), $T_A = 20\ \mu s$ (b), and $T_A = 100\ \mu s$ (c).

We find the minimum energy gaps to be closing exponentially as a function of the problem size in the asymptotic limit for all the three Hamiltonians used for quantum annealing. The scaling exponent for the median minimum energy gap ($r_\Delta$) is observed to be -0.541 for the standard quantum annealing Hamiltonian, whereas $r_\Delta = -0.230$ for the quantum annealing Hamiltonian with the ferromagnetic trigger Hamiltonian, and $r_\Delta = -0.316$ for the quantum annealing Hamiltonian with the antiferromagnetic trigger Hamiltonian. It can therefore be concluded that the addition of the trigger Hamiltonian, especially of the ferromagnetic type, can significantly improve the scaling of the minimum energy gaps.

According to the adiabatic theorem, the scaling exponent of the annealing time required for an adiabatic evolution of the system is approximately twice the scaling exponent of the correlation length (i.e., inverse of the minimum energy gap). Hence, we can expect the median theoretical runtime to diverge exponentially with a rate $r_{TR} \approx 1.082$ for the standard quantum annealing Hamiltonian, which is worse than the scaling of a brute force search of the ground state that scales with an exponent of 0.693. However this scaling is expected to be significantly improved with the addition of the ferromagnetic trigger Hamiltonian ($r_{TR} = 0.460$) and the antiferromagnetic trigger Hamiltonian ($r_{TR} = 0.632$).

The next important observable studied in this work is the success probability, which is defined as the probability of obtaining the ground state of the problem Hamiltonian at the end of the annealing process. We obtained the success probabilities from our quantum annealing simulations using the three Hamiltonians and three annealing times $T_A$ = 10, 100, and 1000. For a comparison of the results obtained from the simulations with those obtained using quantum annealing hardware, we obtain the success probabilities using the systems DW2000Q and DWAdv offered by D-Wave, choosing the annealing times of 4, 20, and 100 $\mu s$.

From the simulation results, we obtain three kinds of success probability distributions, as indicated by mapping the distributions of the minimum energy gaps in the adiabatic limit, for the problem sets corresponding to $N = 16, 17, 18$. The first is the bimodal distribution, which is obtained for the quantum annealing algorithm using the standard Hamiltonian, for all the annealing times, and for the algorithm with the antiferromagnetic trigger Hamiltonian for annealing times of 100 and 1000. The second type of distribution, exhibited by the quantum annealing algorithm using the ferromagnetic trigger Hamiltonian for all the annealing times, is unimodal. Lastly, the success probability distribution for the annealing algorithm with the antiferromagnetic trigger Hamiltonian for the short annealing time of 10 is constant. Unlike [31], we do not find the bimodality of the success probability distribution to be a signature of quantum annealing.

Interestingly, the success probability distributions obtained from the D-Wave systems also show these three kinds of distributions. While the success probability distributions resulting from DW2000Q are bimodal for all annealing times, as was the case for the simulations using the standard quantum annealing Hamiltonian, the distributions obtained from DWAdv are found to be unimodal and constant depending on the choice of the annealing time and the problem set.

TABLE I: (color online) Median scaling exponents $r_\Delta$ and $r_{TTS99}$ for minimum energy gaps and TTS99 for $T_A = 10, 100, 1000$, respectively, obtained from simulations for the three quantum annealing Hamiltonians. The scaling exponents for the theoretical run-times are $r_{TR} = 2|r_\Delta|$.

| Hamiltonian | $r_\Delta$ | $r_{TTS99}$ | | |
|---|---|---|---|---|
| | | $T_A = 10$ | $T_A = 100$ | $T_A = 10$ |
| Standard | -0.541 | 0.530 | 1.170 | 1.205 |
| Ferromagnetic trigger | -0.230 | 0.373 | 0.475 | 0.478 |
| Antiferromagnetic trigger | -0.316 | 0.277 | 0.720 | 0.619 |

TABLE II: (color online) Median scaling exponent $r_{TTS99}$ for $T_A = 4, 20, 100$ $\mu s$, obtained by using DW2000Q and DWAdv.

| Device | $r_{TTS99}$ | | |
|---|---|---|---|
| | $T_A = 4$ $\mu s$ | $T_A = 20$ $\mu s$ | $T_A = 100$ $\mu s$ |
| DW2000Q | 0.674 | 0.511 | 0.531 |
| DWAdv | 0.393 | 0.325 | 0.302 |

However, the similar results from the simulations correspond to quantum annealing Hamiltonians with ferromagnetic and antiferromagnetic triggers, respectively. This suggests, that slightly different mechanisms are at play in the two D-Wave systems and noise and temperature effects are more dominant in DWAdv. A better understanding of these mechanisms and differences calls for further investigation.

The last observable studied in this work is the run-time required to obtain at least one solution, in multiple runs of the annealing algorithm, with 99% certainty, i.e., TTS99. This quantity is calculated using the success probabilities obtained from the simulations as well as the D-Wave systems, for different annealing times.

Focusing first on the simulation results, it is observed that for all the three quantum annealing Hamiltonians studied, the adiabatic theorem can predict the scaling exponent of TTS99 $r_{TTS99}$ based on $r_\Delta$, in the long annealing time regime (see Tab. I). In this case, the addition of the ferromagnetic trigger Hamiltonian to the standard Hamiltonian for quantum annealing, as predicted by the adiabatic theorem, yields the smallest $r_{TTS99}$. The values of $r_{TTS99}$ for the short annealing time of 10, however, are significantly smaller than $r_{TR}$, for the three Hamiltonians. This is the regime where the annealing time is not long enough for the state of the system to evolve adiabatically, and hence different non-adiabatic mechanisms can play an important role in enhancing the success probabilities. Since such mechanisms are enhanced upon adding the antiferromagnetic trigger Hamiltonian to the standard Hamiltonian for quantum annealing, the corresponding algorithm has the smallest value of $r_{TTS99}$ in this case.

A similar analysis of the success probabilities obtained with the D-Wave systems also yields interesting results. It is observed that for both the systems, the value of $r_{TTS99}$ for the median TTS99 is much smaller than the theoretically predicted $r_{TR}$ (see Tab. II). Furthermore, $r_{TTS99}$ becomes smaller as the annealing time is increased on both the systems, except for $T_A = 20$ $\mu s$ obtained with DW2000Q. The $r_{TTS99}$ from DW2000Q for $T_A = 100$ $\mu s$ is comparable to $r_{TTS99}$ obtained from the simulations for $T_A = 10$. Furthermore, the corresponding $r_{TTS99}$ from DWAdv is even smaller for all the annealing times. These observations suggest that the D-Wave systems are not working in the ideal quantum limit, and that noise and temperature effects play a significant role. The observations also confirm that these effects are more prominent in DWAdv. Furthermore, the scaling exponents are closer to those obtained from the simulations in the fast annealing limit than the simulation results for long annealing times.

Lastly, we compare the performance of the quantum annealing algorithm for solving the sets of 2-SAT problems with that of simulated classical annealing. It was found that the median run-time for the latter increases exponentially, with a rate $r_{SA} = 0.34$ [43]. This suggests that for these problems, simulated annealing has a better scaling than quantum annealing using the standard Hamiltonian. However, for the quantum annealing algorithm with the antiferromagnetic trigger Hamiltonian added to the standard Hamiltonian and for a short annealing time, we obtain a better scaling of the median TTS99 ($r_{TTS99} = 0.277$). This is also the case for the TTS99 scaling obtained from DWAdv for long annealing times ($r_{TTS99} = 0.325$ for $T_A = 20$ $\mu s$, and $r_{TTS99} = 0.302$ for $T_A = 100$ $\mu s$). Hence, for an improved performance of the quantum annealing algorithm modifications bolstering the non-adiabatic mechanisms should be introduced to the standard Hamiltonian.

It should be emphasized that there are known classical algorithms that can solve the 2-SAT problems in polynomial time. On the other hand, in order to employ quantum annealing for solving 2-SAT problems, they need to be mapped to the Ising Hamiltonian, which aims at yielding the ground state of the problem Hamiltonian from an exponentially large Hilbert space. One of the possible ways of improving the performance of quantum annealing for solving the 2-SAT problems is thus by incorporating features from the more-efficient classical algorithms, like directed graphs and identifying strongly connected components, in mapping the problem to the Ising model.

## VII. ACKNOWLEDGEMENTS

The authors gratefully acknowledge the Gauss Centre for Supercomputing e.V. (www.gauss-centre.eu) for funding this project by providing computing time through the John von Neumann Institute for Computing (NIC) on the GCS Supercomputer JUWELS [65] at Jülich Supercomputing Centre (JSC). The authors also gratefully acknowledge the computing time granted through JARA on the supercomputer JURECA [66] at Forschungszentrum Jülich. V.M. acknowledges support from the project JUNIQ that has received funding from the German Federal Ministry of Education and Research (BMBF) and the Ministry of Culture and Science of the State of North Rhine-Westphalia.

**Appendix: Distributions**

We present the probability distribution functions used in this work to fit the distributions of the minimum energy gap, or its inverse, the correlation length. The first such distribution is the Fréchet distribution,

$$F_k(x) = a\left(\frac{b}{x}\right)^{k+1} e^{-\left(\frac{b}{x}\right)^k}, \tag{A.1}$$

where $a, b$, and $k$ are constants. The second is the Weibull distribution, which is given by

$$W_k(x) = a\left(\frac{x}{b}\right)^{k-1} e^{-\left(\frac{x}{b}\right)^k}. \tag{A.2}$$

We also define the translated-Weibull distribution,

$$W_k(x) = a\left(\frac{x-\mu}{b}\right)^{k-1} e^{-\left(\frac{x-\mu}{b}\right)^k}, \tag{A.3}$$

where $\mu$ is the extent of the translation. Lastly, we derive the distribution for a variable $y = 1/x$, if the distribution for $x$ is found to follow the translated-Weibull distribution. The Jacobian for this mapping is $J = ||\partial x/\partial y|| = 1/x^2$, and therefore

$$W_k(y) = a(1-\mu y)^2\left(\frac{1-\mu y}{by}\right)^{k+1} e^{-\left(\frac{1-\mu y}{b\Delta_{min}}\right)^k}. \tag{A.4}$$

This distribution will be referred to as the transformed-translated-Weibull distribution.